\documentclass[11pt,twoside]{article}

\usepackage{asp2006}
\usepackage{epsf}
\usepackage{psfig}
\usepackage{lscape}

\begin{document}
\title{The Mass-to-Light Ratios of Galactic Globular Clusters}   
\author{J.~M.~Diederik Kruijssen$^{1,2}$ and Steffen Mieske$^3$}   
\affil{$^1$Astronomical Institute, Utrecht University, PO Box 80000, 3508 TA Utrecht, The Netherlands; {\tt kruijssen@astro.uu.nl}}
\affil{$^2$Leiden Observatory, Leiden University, PO Box 9513, 2300 RA Leiden, The Netherlands}
\affil{$^3$European Southern Observatory, Alonso de Cordova 3107, Vitacura, Santiago, Chile}
\begin{abstract} 
The observed mass-to-light ($M/L$) ratios of globular clusters (GCs) are on average $\sim 20\%$ lower than expected from Simple Stellar Population (SSP) models, which only account for the effects of stellar evolution. We study the $M/L$ ratio evolution of a sample of 24 Galactic GCs using parameterised cluster models. The dynamical evolution of GCs is included by accounting for their dissolution and by using a detailed description of the evolution of the stellar mass function. The ejection of low-mass stars leads to a decrease of $M/L$, which is found to explain the discrepancy between the observations and SSP models.
\end{abstract}



\section{Introduction}
Galaxy mergers are thought to have been the formation sites of globular cluster systems (GCSs). As such, globular clusters (GCs) can be used as tracers for studying the early evolution of galaxies and any mergers these galaxies may have experienced. To relate the GCS of a galaxy to its formation history, it is essential to obtain a good consensus on the present masses of the GCs. For obtaining these, (constant) mass-to-light ($M/L$) ratios are commonly used \citep{fall01,jordan07,mclaughlin08}. However, the observed dynamical $M/L$ ratios of GCs in several galaxies are found to be $\sim20\%$ lower than the values expected from Simple Stellar Population (SSP) models \citep{mandushev91,mclaughlin05,rejkuba07}. This complicates the interpretation of GC masses, as it first needs to be understood why their $M/L$ ratios deviate from those predicted by SSP models\footnote{Interestingly, ultra-compact dwarf galaxies (UCDs), which have been proposed to represent a continuation of the GC mass range to higher masses, have $M/L$ ratios that are $\sim 25\%$ {\it higher} than those predicted by SSP models \citep{rejkuba07,mieske08}. We separate this discrepancy from the one between GCs and SSP models and only consider GCs.}.

It has been proposed that the $M/L$ ratio difference is due to the dynamical evolution of GCs \citep{kruijssen08,kruijssen08b}. The preferential ejection of low-mass, high-$M/L$ stars from dissolving star clusters changes the shape of the stellar mass function (MF) within a cluster \citep{vesperini97b,baumgardt03,demarchi07}, causing the luminosity to be only marginally affected while the mass decreases. As such, dissolution decreases the $M/L$ ratio with respect to the expected evolution from SSP models, in which it is assumed that the shape of the stellar MF does not vary.

In \citet{kruijssen09}, the hypothesis of low-mass star depletion as an explanation for the low $M/L$ ratios of GCs was tested by considering the sample of 24 Galactic GCs for which the orbits \citep{dinescu99} and observed dynamical $M/L$ ratios \citep{mclaughlin05} are known. There, we derived the dissolution timescales from the individual cluster orbits and computed the resulting $M/L$ evolution for the GCs in the sample. It was found that low-mass star depletion can indeed account for the $\sim 20\%$ gap between the observations and SSP models. As a first-order approximation, low-mass star depletion was included by increasing the lower stellar mass limit in the cluster, which was tuned to $N$-body simulations in order to give reliable results. In reality, the stellar MF evolves more gradually. This has been included in new, physical models of the evolution of the stellar MF in dissolving star clusters \citep{kruijssen09c}. The aim of the present paper is to revisit the calculations of \citet{kruijssen09} with these new models and to verify whether their conclusions still hold.

\section{Star cluster evolution and mass-to-light ratio}
We use the parameterised cluster model {\tt SPACE} \citep{kruijssen08b,kruijssen09c}, which incorporates the effects of stellar evolution, stellar remnant production, dynamical dissolution and energy equipartition. The dissolution timescale due to evaporation \citep[cf.][]{baumgardt03} and tidal shocks \citep[cf.][]{dinescu99} is determined for each of the 24 GCs in our sample by considering their individual orbits \citep{kruijssen09} and is subsequently converted to a mass loss rate \citep{lamers05}.

The evolution of the stellar MF is computed by considering the ejection rate as a function of stellar mass \citep{kruijssen09c}. The adopted method accounts for mass segregation and dissolution in a tidal field by using the timescale on which energy equipartition is reached for different stellar masses and by comparing the stellar velocities with the escape velocity. The photometry is computed by integrating the MF over the new Padova isochrones \citep{marigo08}.

\section{Comparison to observations}
\begin{figure}
\plottwo{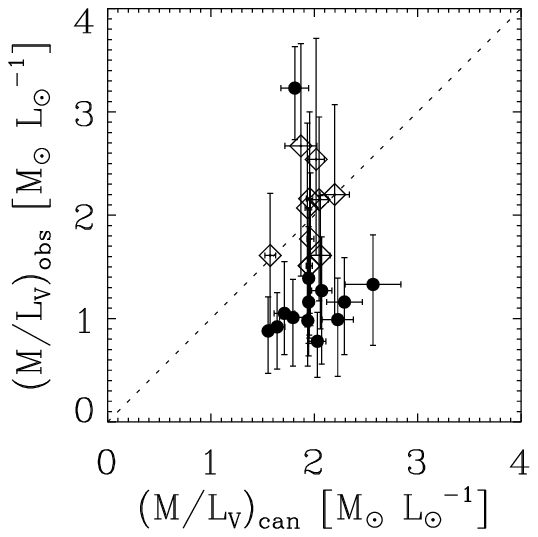}{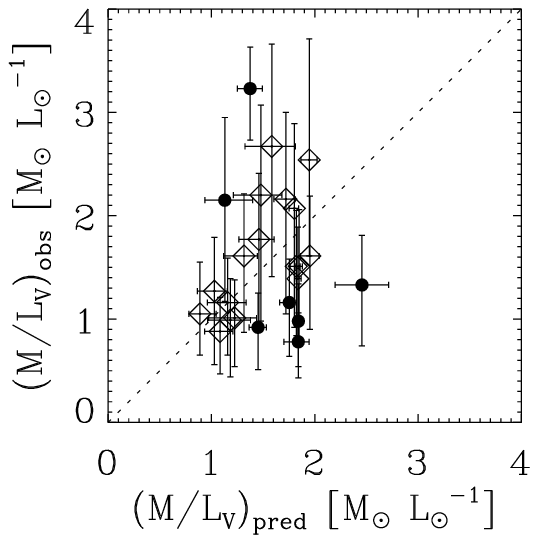}
\caption[]{\label{fig:ml}
    Comparison of $V$-band $M/L$ ratios that are predicted by SSP models (left) with observed $M/L_V$ ratios, and of our modeled $M/L_V$ ratios (right) with the observed values. Error bars denote 1$\sigma$ standard errors, with diamonds marking GCs for which theory and observation are in $1\sigma$ agreement, and dots representing the remaining deviant GCs.
    }
\end{figure}
The $V$-band $M/L$ ratios from SSP models are compared to the observed values in the left-hand panel of Fig.~\ref{fig:ml}. In the right-hand panel, the modeled $V$-band $M/L$ ratios for the 24 GCs in our sample are compared to the observations. The improvement with respect to the left-hand panel is considerable, but there remains a number of GCs with observed $M/L$ ratios that are lower than the modeled ones. This was explained by \citet{kruijssen09}, who reasoned that some of the observed $M/L$ ratios from \citet{mclaughlin05} are likely biased due to their use of isotropic single-mass King models to determine the $M/L$ ratios. This method leads to a tendency towards the central $M/L$ ratio rather than the global one for GCs with evolved internal structure. 

\begin{figure}
\center\resizebox{7.3cm}{!}{\plotone{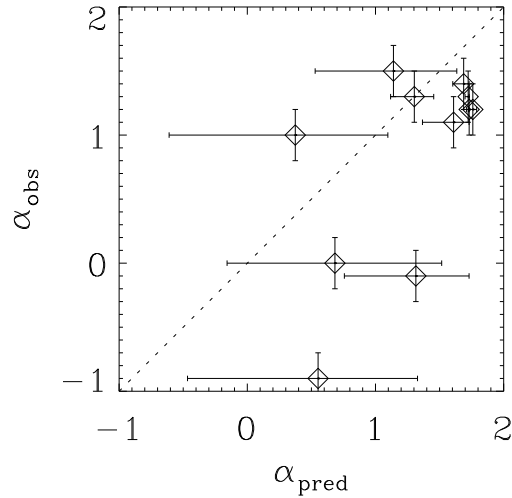}}
\caption[]{\label{fig:mf}
    Observed stellar MF slope $\alpha_{\rm obs}$ for the stellar mass range $0.3<m/{\rm M}_\odot<0.8$ versus the modeled mean slope in that range $\alpha_{\rm pred}$. Error bars are 1$\sigma$ standard errors, mainly caused by the uncertainty in the GC mass.
    }
\end{figure}

In total, our new models explain the $M/L$ ratios of 17 out of 24 GCs within the $1\sigma$ error margins. This is a substantial development with respect to \citet{kruijssen09}, where half of the GCs was explained. The use of the new MF models and of the new Padova isochrones contribute equally to this improvement (three GCs each). The average fraction of the observed $M/L_V$ ratio with respect to the value predicted by SSP models is 0.81$^{+0.06}_{-0.08}$, while for our modeled fraction this is 0.79$\pm 0.01$. These values are in excellent agreement. An independent check of our models is provided by comparing the modeled stellar MF slopes to those observed by \citet{demarchi07}. This is shown in Fig.~\ref{fig:mf} and confirms the general validity of our models, despite the scatter and large error bars.

\section{Discussion}
Low-mass star depletion in dissolving GCs explains the $\sim 20\%$ discrepancy between the observed $M/L$ ratios and those predicted by SSP models. The new cluster models \citep{kruijssen09c}, in which we account for the changing slope of the stellar MF rather than shifting the lower stellar mass limit, show that the results from \citet{kruijssen09} also hold when more detailed models are applied, and improve the agreement between theory and observations.

Nonetheless, care should be taken when using observed dynamical $M/L$ ratios that neglect a stellar mass spectrum or the variability of $M/L$ from the centre to the outskirts of a cluster. These cannot be interpreted as global $M/L$ ratios if the $M/L$ too strongly varies of radius. Considering the central role GCs play in (extra)galactic astronomy, it essential to obtain an accurate census of their dynamical masses. This would improve the observed globular cluster mass function \citep{kruijssen09b}, which can then be more accurately used to trace and interpret the formation history of galaxies.

\acknowledgements 
JMDK thanks Henny Lamers for advice, and is grateful to East Tennessee State University for hosting an excellent conference and for financially supporting attendance. The Leids Kerkhoven-Bosscha Fonds is acknowledged for supporting attendance to the conference. JMDK is supported by a TopTalent fellowship from the Netherlands Organisation for ScientiÞc Research (NWO), grant number 021.001.038.



\end{document}